\documentclass[11 pt]{article}

\usepackage{graphicx}
\usepackage{latexsym}
\usepackage{amsfonts}
\usepackage{amsmath}
\usepackage{enumerate}
\usepackage{amssymb}
\usepackage{ifpdf}

\ifpdf
\fi

\newcommand\blackslug{\hbox{\hskip 1pt \vrule width 4pt height 8pt depth 1.5pt
        \hskip 1pt}}
\newcommand\bbox{\hfill \quad \blackslug \medbreak}
\def\d{\hbox{-}}
\def\c{\hbox{-}\cdots\hbox{-}}

\newtheorem{theorem}{Theorem}[section]
\newtheorem{lemma}[theorem]{Lemma}
\newtheorem{question}[theorem]{Question}
\newtheorem{conjecture}[theorem]{Conjecture}

\newcounter{claim}

\newcommand{\Proof}{\setcounter{claim}{0}\noindent{\bf Proof.}\ \ }

\title{A polynomial turing-kernel for weighted independent set in bull-free
  graphs} \author{St\'ephan Thomass\'e\thanks{CNRS, LIP, ENS
    de Lyon, INRIA, Universit\'e de Lyon.
    Partially supported by ANR project Stint under reference ANR-13-BS02-0007}~, Nicolas
  Trotignon\footnotemark[1]~ and Kristina Vu\v skovi\'c\thanks{School of
    Computing, University of Leeds, Leeds LS2 9JT, UK; and Faculty of Computer Science
    (RAF), Union University, Knez Mihajlova 6/VI, 11000 Belgrade, Serbia.  Partially supported by
    EPSRC grant EP/K016423/1 and Serbian Ministry of Education and
    Science projects 174033 and III44006.}}

\begin{document}
\maketitle

\begin{abstract}
  The maximum stable set problem is NP-hard, even when restricted to
  triangle-free graphs. In particular, one cannot expect
  a polynomial time algorithm deciding if a bull-free graph has a
  stable set of size $k$, when $k$ is part of the instance. Our main
  result in this paper is to show the existence of an FPT algorithm
  when we parameterize the problem by the solution size $k$.  A
  polynomial kernel is unlikely to exist for this problem. We show
  however that our problem has a polynomial size Turing-kernel.  More
  precisely, the hard cases are instances of size ${O}(k^5)$.  As a
  byproduct, if we forbid odd holes in addition to the bull, we show
  the existence of a polynomial time algorithm for the stable set
  problem.  We also prove that the chromatic number of a bull-free
  graph is bounded by a function of its clique number and the maximum
  chromatic number of its triangle-free induced subgraphs.  All our
  results rely on a decomposition theorem for bull-free graphs due to
  Chudnovsky which is modified here, allowing us to provide extreme
  decompositions, adapted to our computational purpose.
\end{abstract}

\section{Introduction}

In this paper all graphs are simple and finite.  We say that a graph
$G$ {\em contains} a graph $F$, if $F$ is isomorphic to an induced
subgraph of $G$. We say that $G$ is {\em $F$-free} if $G$ does not
contain $F$. For a class of graphs ${\cal F}$, the graph $G$ is {\em
  ${\cal F}$-free} if $G$ is $F$-free for every $F \in {\cal F}$.  The
{\em bull} is a graph with vertex set $\{ x_1,x_2,x_3,y,z\}$ and edge
set $\{ x_1x_2,x_1x_3,x_2x_3,x_1y,x_2z\}$.  A \emph{hole} in a graph
is an induced subgraph isomorphic to a chordless cycle of length at
least~4.  A hole is \emph{odd} or \emph{even} according to the parity
of its number of  vertices.

\vspace{2ex}

Chudnovsky in a series of papers \cite{bull1, bull2, bull3, bull4}
gives a complete structural characterisation of bull-free graphs (more
precisely, bull-free trigraphs, where a trigraph is a graph with some
adjacencies left undecided).  Roughly speaking, this theorem asserts that every
bull-free trigraph is either in a well-understood \emph{basic} class,
or admits a \emph{decomposition} allowing to break the trigraph into
smaller \emph{blocks}.  In Section~\ref{sec:def}, we extract what we
need for the present work, from the very complex theorem of
Chudnovsky.

\vspace{2ex}

In Section~\ref{sec:extDec}, we prove that bull-free trigraphs admit
\emph{extreme} decompositions, that are decompositions such that one
of the blocks is basic.  It is very convenient for design of fast
algorithms and proofs by induction.

\vspace{2ex}

In Section~\ref{sec:algoDec}, we give polynomial time algorithms to
actually compute the extreme decompositions whose existence is proved
in the previous section.

\vspace{2ex}

In Section~\ref{sec:bofFree}, we apply the previous results to give a
polynomial time algorithm that computes $\alpha(G)$ in any \{bull, odd
hole\}-free graph, where $\alpha(G)$ denotes the maximum size of an
{\em independent set} (or {\em stable set}) of a graph $G$, that is a
subset of the vertex-set of $G$ no two vertices of which are
adjacent.  We also solve the weighted version of this problem and our algorithm
is \emph{robust}, meaning that it can be run on any graph and either
outputs the correct answer or a certificate showing that the graph is
not in the class.  
 This result is known already.  Brandst{\"a}dt and
 Mosca~\cite{brandstadtMo:2012}
gave a more direct algorithm for the same problem.  We present our
algorithm because it illustrates well our method to compute $\alpha$
(and most of the material of Section~\ref{sec:bofFree} is needed in
the rest of the paper). 

Note that computing $\alpha$ is NP-hard in general, and it remains
difficult even when seemingly a lot of structure is imposed on the
input graph. For example, it remains NP-hard for triangle-free
graphs~\cite{poljak:74}, and hence for bull-free graphs.  The
complexity of computing $\alpha$ and $\chi$ in odd-hole-free graphs is
not known.

\vspace{2ex}

In Section~\ref{sec:decAlpha}, we give an FPT-algorithm for the
maximum stable set problem restricted to bull-free graphs.  Let us
explain this.  The notion of fixed-parameter tractability (FPT) is a
relaxation of classical polynomial time solvability.  A parameterized
problem is said to be {\em fixed-parameter tractable} if it can be
solved in time $f(k)P(n)$ on instances of input size $n$, where $f$ is
a computable function (so $f(k)$ depends only on the value of
parameter $k$), and $P$ is a polynomial function independent of $k$.
We give an FPT-algorithm for the maximum stable set problem restricted
to bull-free graphs.  This generalizes the result of Dabrowski, Lozin,
M\"{u}ller and Rautenbach~\cite{dlmr} who give an FPT-algorithm for
the same parameterized problem for \{bull, $\overline{P_5}$\}-free
graphs, where $P_5$ is a path on 5 vertices and $\overline{P_5}$ is
its complement.  In a weighted graph the {\em weight} of a set is the
sum of the weights of its elements, and with $\alpha_w(G)$ we denote
the weight of a maximum weighted independent set of a graph $G$ with
weight function $w$.  We state below the problem that we solve more
formally.

\vspace{2ex}

\noindent
{\sc parameterized weighted independent set}
\\
{\em Instance:} A weighted graph $G$ with weight function
$w:V(G)\longrightarrow \mathbb{N}$ and a positive integer $k$.
\\
{\em Parameter:} $k$
\\
{\em Problem:} Decide whether $G$ has an independent set of weight at
least $k$.  If no such set exists, find an independent set of weight
$\alpha _w(G)$.

\vspace{2ex} 

Observe that the problem above is hard for general graphs.
Furthermore, it is $W[1]$-hard~\cite{df}.  

\vspace{2ex}

In Section~\ref{sec:polyOracle}, we show that while a polynomial
kernel is unlikely to exist since the problem is OR-compositional, we
can prove nonetheless that the hardness of the problem can be reduced
to polynomial size instances. Precisely we show that if it takes time
$f(k)$ to decide if a stable set of size $k$ exists for bull-free
graphs of size ${O}(k^5)$, then one can solve the problem on instances
of size $n$ in time $f(k)P(n)$ for some polynomial $P$ in $n$. The
fact that hard cases can be reduced to size polynomial in $k$ is not
captured by the existence of a polynomial kernel, but by what is
called a Turing-kernel (see Section~\ref{sec:polyOracle} or
Lokshtanov~\cite{lokshtanov:phd} for a definition of Turing-kernels).
Even the existence of a Poly($n$) set of kernels of size Poly($k$)
seems unclear for this problem. To our knowledge, stability in
bull-free graphs is the first example of a problem admitting a
polynomial Turing-kernel which is not known to have an independent set
of polynomial kernels.  Further examples are given in
Jansen~\cite{jansen:2014}.  An interesting question is to investigate
which classical problems without polynomial kernels do have a
polynomial Turing-kernel. This question is investigated by Hermelin et
al.~\cite{hermelin}.

\vspace{2ex}

All this work has been very recently improved by Perret du Cray and
Sau~\cite{PerretSau:14}.  Using the same method as ours, but with a
better implementation for detecting the decomposition and a more
precise description of the basic classes, they reach a running time of
$2^{O(k^2)} n^7$, and they could get the size of the Turing kernel
down to $O(k^2)$.

\vspace{2ex}

At the end of the paper, we use the machinery developped in the
previous sections to bound the chromatic number of bull-free graphs.
Let $\chi(G)$ denote the chromatic number of $G$ and $\omega(G)$
denote the maximum size of a {\em clique} of a graph $G$, that is a
set of pairwise adjacent vertices of $G$.  An obvious reason for a
graph to have a high chromatic number is the presence of large
cliques.  But as shown by many well-known constructions, this is not
the only source: there exist graphs with fixed maximum clique size,
namely~2, and arbitrarily large chromatic number.  Therefore, a second
reason for a graph to have a large chromatic number can be the presence of
triangle-free induced subgraphs with large chromatic number.  We therefore define the
\emph{triangle-free chromatic number} of a graph $G$ as the maximum
chromatic number of a triangle-free induced subgraph of $G$, and we denote it
by $\chi_T(G)$.  We wonder whether the only possible reason why
a graph $G$ may have a large chromatic number is that $\chi_T(G)$ is
large or $\omega(G)$ is large.  This has been asked several times by
researchers, but we could not find a reference. It can be stated formally as follows.

\begin{question}
  \label{q:fb}
Does there exits a function $f$ such that for every graph $G$ $$ \chi(G)
\leq f(\chi_T(G), \omega(G)) $$
\end{question}

Note that if we forget the word ``induced'' in the definition of
$\chi_T$, the function exists as shown by R\"odl~\cite{rodl:76}.  In
Section~\ref{sec:bChi}, we prove the existence of $f$ for bull-free
graphs.  The existence of $f$ in general would maybe not be so
surprising, since with respect to the chromatic number, triangle-free
graph are perhaps as complex as general graphs.  Nevertheless, it
would have non-trivial implications, in particular it would settle the
famous conjecture below on odd-hole-free graph.  A class of graphs is
\emph{hereditary} if it is closed under taking induced subgraphs.  It
is $\chi$-bounded if there exists a function $f$ such every graph $G$
of the class satisfies $\chi(G) \leq f(\omega(G))$.

\begin{conjecture}[Gy\'arf\'as~\cite{gyarfas:perfect}]
 The class of odd-hole-free graph is $\chi$-bounded.
\end{conjecture}

A ``yes'' answer to Question~\ref{q:fb} would settle the conjecture
above because, $\chi_T$ is at most 2 in graphs with no odd holes.
Indeed, an odd-hole-free graph with no triangle is bipartite.  Our
results imply that Gy\'arf\'as's conjecture on odd holes is true for
bull-free graphs.

\vspace{2ex}

To conclude, it seems to us that the existence of a polynomial time
algorithm for the maximum stable set problem for a class of graphs is
such a strong property that it raises the next question.  

\begin{question}
  Is it true that if a polynomial time algorithm exists for the
  maximum stable set problem in a hereditary class of graph, then this
  class is $\chi$-bounded?
\end{question}

If P=NP, then the above question is clearly answered by ``no'' (since
the class of all graphs is not $\chi$-bounded); but under the
assumption that P$\neq$NP, it might be answered by ``yes''.  Note that
to our knowledge, all hereditary classes with a polynomial time
algorithm for $\alpha$ are $\chi$-bounded, and in some situations, the
algorithm for $\alpha$ is quite involved (for instance in perfect
graphs~\cite{gls}, claw-free
graphs~\cite{faenzaOrioloStauffer:clawFree}, or in $P_5$-free graphs
as announced recently by Lokshtanov, Vatshelle and
Villanger~\cite{loVaVi:13}).

\vspace{2ex}

A shorter version of the present work appeared in
\cite{thomasseTrVu:bf}.

\section{Decomposition of bull-free graphs}
\label{sec:def}

In the series of papers \cite{bull1,bull2,bull3,bull4} Chudnovsky
gives a complete structural characterisation of bull-free graphs which
we first describe informally.  Her construction of all bull-free
graphs starts from three explicitly constructed classes of basic
bull-free graphs: ${\cal T}_0,{\cal T}_1$ and ${\cal T}_2$.  Class
${\cal T}_0$ consists of graphs whose size is bounded by some constant,
the graphs in ${\cal T}_1$ are built from a triangle-free graph $F$
and a collection of disjoint cliques with prescribed attachments in
$F$ (so triangle-free graphs are in this class, and also ordered split
graphs), and ${\cal T}_2$ generalizes graphs $G$ that have a pair $uv$
of vertices, so that $uv$ is dominating both in $G$ and $\bar{G}$.
Furthermore, each graph $G$ in ${\cal T}_1 \cup {\cal T}_2$ comes with
a list ${\cal L}_G$ of ``expandable edges''.  Chudnovsky shows that
every bull-free graph that is not obtained by substitution (a
composition operation that is a reversal of homogeneous set
decomposition) from smaller ones, can be constructed from a basic
bull-free graph by expanding the edges in ${\cal L}_G$ (where edge
expansion is an operation corresponding to reversing the homogeneous
pair decomposition).  All these terms will be defined later in this
section.  To prove and use this result, it is convenient to work on trigraphs
(a generalization of graphs where some edges are left ``undecided''),
and the first step is to obtain a decomposition theorem for bull-free
trigraphs using homogeneous sets and homogeneous pairs.  In this paper
we  need a simplified statement of this decomposition theorem,
which we now describe formally.

\subsection*{Trigraphs}

For a set $X$, we denote by $X \choose 2$ the set of all subsets of
$X$ of size~2. For brevity of notation an element $\{ u,v \}$ of $X
\choose 2$ is also denoted by $uv$ or $vu$. A {\em trigraph} $T$
consists of a finite set $V(T)$, called the {\em vertex set} of $T$,
and a map $\theta : {{V(T)} \choose 2} \longrightarrow \{ -1,0,1 \}$,
called the {\em adjacency function}.

Two distinct vertices of $T$ are said to be {\em strongly adjacent} if
$\theta(uv)=1$, {\em strongly antiadjacent} if $\theta(uv)=-1$, and
{\em semiadjacent} if $\theta(uv)=0$. We say that $u$ and $v$ are
{\em adjacent} if they are either strongly adjacent, or semiadjacent;
and {\em antiadjacent} if they are either strongly antiadjacent, or
semiadjacent. An \emph{edge} (\emph{antiedge}) is a pair of adjacent
(antiadjacent) vertices. If $u$ and $v$ are adjacent (antiadjacent),
we also say that $u$ is {\em adjacent (antiadjacent) to} $v$, or that
$u$ is a {\em neighbor (antineighbor)} of $v$. Similarly, if $u$ and
$v$ are strongly adjacent (strongly antiadjacent), then $u$ is a {\em
  strong neighbor (strong antineighbor)} of $v$. 

Let $\eta(T)$ be the set of all strongly adjacent pairs of $T$,
$\nu(T)$ the set of all strongly antiadjacent pairs of $T$, and
$\sigma(T)$ the set of all semiadjacent pairs of $T$. Thus, a trigraph
$T$ is a graph if $\sigma(T)$ is empty. A pair $\{u, v\} \subseteq
V(T)$ of distinct vertices is a \emph{switchable pair} if $\theta(uv)
= 0$, a \emph{strong edge} if $\theta(uv) = 1$ and a \emph{strong
  antiedge} if $\theta(uv) = -1$.  An edge $uv$ (antiedge, strong
edge, strong antiedge, switchable pair) is \emph{between} two sets $A
\subseteq V(T)$ and $B \subseteq V(T)$ if $u\in A$ and $v \in B$, or
if $u \in B$ and $v \in A$.

The \emph{complement} $\overline{T}$ of $T$ is a
trigraph with the same vertex set as $T$, and adjacency function
$\overline{\theta}=-\theta$. 

For $v \in V(T)$, $N(v)$ denotes the set of all vertices in $V(T)
\setminus \{v\}$ that are adjacent to $v$; $\eta(v)$ denotes the set
of all vertices in $V(T) \setminus \{v\}$ that are strongly adjacent
to $v$; $\nu(v)$ denotes the set of all vertices in $V(T) \setminus
\{v\}$ that are strongly antiadjacent to $v$; and $\sigma(v)$ denotes
the set of all vertices in $V(T) \setminus \{v\}$ that are
semiadjacent to $v$.

Let $A \subset V(T)$ and $b \in V(T) \setminus A$. We say that $b$ is
{\em strongly complete} to $A$ if $b$ is strongly adjacent to every
vertex of $A$; $b$ is {\em strongly anticomplete} to $A$ if $b$ is
strongly antiadjacent to every vertex of $A$; $b$ is {\em complete} to
$A$ if $b$ is adjacent to every vertex of $A$; and $b$ is {\em
  anticomplete} to $A$ if $b$ is antiadjacent to every vertex of
$A$. For two disjoint subsets $A,B$ of $V(T)$, $B$ is {\em strongly
  complete (strongly anticomplete, complete, anticomplete)} to $A$ if
every vertex of $B$ is strongly complete (strongly anticomplete,
complete, anticomplete) to $A$. A set of vertices $X\subseteq V(T)$
\emph{dominates (strongly dominates)} $T$ if for all $v\in
V(T)\setminus X$, there exists $u\in X$ such that $v$ is adjacent
(strongly adjacent) to $u$.

A {\em clique} in $T$ is a set of vertices all pairwise adjacent, and
a {\em strong clique} is a set of vertices all pairwise strongly
adjacent. A {\em stable set} is a set of vertices all pairwise
antiadjacent, and a {\em strongly stable set} is a set of vertices all
pairwise strongly antiadjacent. For $X \subset V(T)$ the trigraph
{\em induced by $T$ on $X$} (denoted by $T[X]$) has vertex set $X$,
and adjacency function that is the restriction of $\theta$ to $X
\choose 2$. Isomorphism between trigraphs is defined in the natural
way, and for two trigraphs $T$ and $H$ we say that $H$ is an {\em
 induced subtrigraph} of $T$ (or $T$ {\em contains $H$ as an induced
 subtrigraph}) if $H$ is isomorphic to $T[X]$ for some $X \subseteq
V(T)$. Since in this paper we are only concerned with the induced subtrigraph
containment relation, we say that \emph{$T$ contains~$H$} if $T$
contains $H$ as an induced subtrigraph. We denote by $T\setminus X$
the trigraph $T[V(T) \setminus X]$.

Let $T$ be a trigraph. A \emph{path} $P$ of $T$ is a sequence of
distinct vertices $p_1, \dots, p_k$ such that $k\geq 1$ and for $i,
j \in \{1, \ldots, k\}$, $p_i$ is adjacent to $p_j$ if $|i-j|=1$ and
$p_i$ is antiadjacent to $p_j$ if $|i-j|>1$. Under these
circumstances, $V(P) = \{p_1, \dots, p_k\}$ and we say that $P$ is a
path {\em from $p_1$ to $p_k$}, its {\em interior} is the set
$P^*=V(P) \setminus \{p_1,p_k\}$, and the {\em length} of $P$ is
$k-1$. We also say that $P$ is a \emph{$(k-1)$-edge-path}. Sometimes,
we denote $P$ by $p_1 \c p_k$.  Observe that, since a graph is also a
trigraph, it follows that a path in a graph, the way we have defined
it, is what is sometimes in literature called a chordless path.

A {\em hole} in a trigraph $T$ is an induced subtrigraph $H$ of $T$
with vertices $h_1, \ldots, h_k $ such that $k \geq 4$, and for $i,j
\in \{1, \ldots, k\}$, $h_i$ is adjacent to $h_j$ if $|i-j|=1$ or
$|i-j|=k-1$; and $h_i$ is antiadjacent to $h_j$ if $1<|i-j|<k-1$. The
{\em length} of a hole is the number of vertices in it. Sometimes we
denote $H$ by $h_1 \c h_k \d h_1$. An {\em antipath} ({\em antihole})
in $T$ is an induced subtrigraph of $T$ whose complement is a path
(hole) in $\overline{T}$.

A {\em semirealization} of a trigraph $T$ is any trigraph $T'$ with
vertex set $V(T)$ that satisfies the following: for all $uv \in
{{V(T)} \choose 2}$, if $uv \in \eta(T)$ then $uv \in \eta(T')$, and
if $uv \in \nu(T)$ then $uv \in \nu(T')$.  Sometimes we will describe
a semirealization of $T$ as an {\em assignment of values} to
switchable pairs of $T$, with three possible values: ``strong edge'',
``strong antiedge'' and ``switchable pair''.  A {\em realization} of
$T$ is any graph that is semirealization of $T$ (so, any
semirealization where all switchable pairs are assigned the value
``strong edge'' or ``strong antiedge'').  For $S \subseteq \sigma
(T)$, we denote by $G^T_S$ the realization of $T$ with edge set $\eta
(T) \cup S$,  so in $G_{S}^T$ the switchable pairs in $S$ are assigned
the value ``edge'', and those in $\sigma(T) \setminus S$ the value
``antiedge''. The realization $G^T_{\sigma(T)}$ is called the {\em
  full realization} of~$T$.

A {\em bull} is a trigraph with vertex set $\{ x_1,x_2,x_3,y,z\}$ such
that $x_1,x_2,x_3$ are pairwise adjacent, $y$ is adjacent to $x_1$ and
antiadjacent to $x_2,x_3,z$, and $z$ is adjacent to $x_2$ and
antiadjacent to $x_1,x_3$.  For a trigraph $T$, a subset $X$ of $V(T)$
is said to be a bull if $T[X]$ is a bull.  A trigraph is {\em
  bull-free} if no induced subtrigraph of it is a bull, or
equivalently, no subset of its vertex set is a bull.

Observe that we have two notions of bulls: bulls as graphs (defined in
the introduction), and bulls as trigraphs.  A bull as a graph can be
seen as a bull as a trigraph.  Also, a trigraph is a bull if
and only if at least one of its realization is a bull (as a graph).
Hence, a trigraph is bull-free if and only if all its realizations are
bull-free graphs.  The complement of a bull is a bull (with both
notions), and therefore, if $T$ is bull-free trigraph (or graph), then
so is $\overline{T}$.

A trigraph $T$ is {\em Berge} if it contains no odd hole and no odd antihole.
Therefore, a trigraph is Berge if and only if its complement is Berge. We observe that
$T$ is Berge if and only if every realization (semirealization) of $T$ is Berge.

\subsection*{Decomposition theorem}

A trigraph is called {\em monogamous} if every vertex of it belongs to
at most one switchable pair (so the switchable pairs form a
matching).  We now state the decomposition theorem for bull-free
monogamous trigraphs.  We begin with the description of the cutsets.

\begin{figure}
  \begin{center}
  \includegraphics{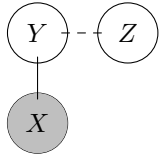}
  \end{center}
  \caption{A homogeneous set.\label{fig:hs}}
\end{figure}

Let $T$ be a trigraph.  A set $X\subseteq V(T)$ is a \emph{homogeneous
  set} in $T$ if $1<|X|<|V(T)|$, and every vertex of $V(T) \setminus
X$ is either strongly complete or strongly anticomplete to $X$.  See
Figure~\ref{fig:hs} (a line means all possible strong edges between
two sets, nothing means all possible strong antiedges, and a dashed
line means no restriction). 

\begin{figure}
  \begin{center}
  \includegraphics{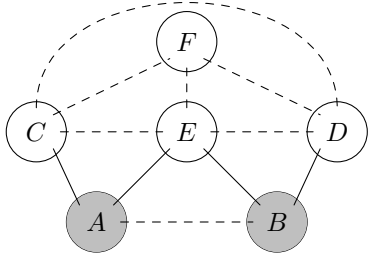}
  \end{center}
  \caption{A homogeneous pair.\label{fig:hp}}
\end{figure}

A {\em homogeneous pair} (see
Figure~\ref{fig:hp}) is a pair of disjoint nonempty subsets
$(A,B)$ of $V(T)$, such that there are disjoint (possibly empty)
subsets $C, D, E, F$ of $V(T)$ whose union is $V(T)\setminus (A\cup
B)$, and the following hold:
\begin{itemize}
\item $A$ is strongly complete to $C \cup E$ and strongly anticomplete to $D\cup F$;
\item $B$ is strongly complete to $D \cup E$ and strongly anticomplete to $C\cup F$;
\item $A$ is not strongly complete and not strongly anticomplete to $B$;
\item $|A\cup B|\geq 3$; and
\item $|C \cup D\cup E\cup F|\geq 3$.
\end{itemize}

Note that ``$A$ is not strongly complete and not strongly anticomplete
to $B$'' does not imply that $|A\cup B|\geq 3$, because it could be
that the unique vertex in $A$ is linked to the unique vertex in $B$ by
a switchable pair.  In these circumstances, we say that $(A,
B,C,D,E,F)$ is a \emph{split} for the homogeneous pair $(A,B)$.  A
homogeneous pair $(A,B)$ is {\em small} if $|A \cup B|\leq 6$.  A
homogeneous pair $(A,B)$ with split $(A,B,C,D,E,F)$ is {\em proper} if
$C\neq \emptyset$ and $D\neq \emptyset$.

We now describe the basic classes.  A trigraph is a \emph{triangle} if
it has exactly three vertices, and these vertices are pairwise
adjacent.  Let ${\cal T}_0$ be the class of all monogamous trigraphs
on at most 8 vertices.  Let ${\cal T}_1$ be the class of monogamous
trigraphs $T$ whose vertex set can be partitioned into (possibly
empty) sets $X, K_1, \ldots ,K_t$ so that $T[X]$ is triangle-free, and
$K_1, \ldots ,K_t$ are strong cliques that are pairwise strongly
anticomplete.  Furthermore, for every $v \in \cup_{i=1}^t K_i$, the
set of neighbors of $v$ in $X$ partitions into strong stable sets $A$
and $B$ such that $A$ is strongly complete to $B$.  In Chudnovsky's
work, the trigraphs in ${\cal T}_0$ are precisely defined, and the
adjacencies between the cliques and $X$ in trigraphs from ${\cal T}_1$
are precisely specified. Furthermore, the homogeneous pairs used are
more structured (in order to allow for the reversal of homogeneous
pair decomposition to be class-preserving), which also leads to the
need for another basic class ${\cal T}_2$.  In our algorithm we do not
need the homogeneous pairs to be so particularly structured, so the
following statement will suffice.  Let $\overline{{\cal T}_1}=\{
\overline{T}:T\in {\cal T}_1\}$.  In 5.6 and 5.7 of~\cite{bull2} it is
shown that if $T$ is a bull-free monogamous trigraph then either $T\in
{\cal T}_0 \cup {\cal T}_1 \cup \overline{{\cal T}_1} \cup {\cal T}_2
\cup \overline{{\cal T}_2}$, or it has a homogeneous set or
homogeneous pair of type 0, 1 or 2. From the definition of these types
of homogeneous pairs it clearly follows that type 0 is a small
homogeneous pair, and type 1 and 2 are proper homogeneous pairs.  From
the definition of class ${\cal T}_2$ it clearly follows that if $T\in
({\cal T}_2 \cup \overline{{\cal T}_2}) \setminus ({\cal T}_0 \cup
{\cal T}_1 \cup \overline{{\cal T}_1})$ then $T$ has a proper
homogeneous pair.  A trigraph is {\em basic} if it belongs to ${\cal
  T}_0\cup {\cal T}_1 \cup \overline{{\cal T}_1}$.  All this implies
the following theorem.

\begin{theorem}[Chudnovsky \cite{bull1,bull2,bull3,bull4}]
\label{thm:bull}
If $T$ is a bull-free monogamous trigraph, then one of the following holds:
\begin{itemize}
\item $T$ is basic;
\item $T$ has a homogeneous set; 
\item  $T$ has a small homogeneous pair; or
\item $T$ has a proper homogeneous pair.
\end{itemize}
\end{theorem}

We do not know whether the theorem above is algorithmic. Deciding
whether a graph is bull-free can clearly be done in polynomial time.
Also, detecting the decompositions is easy (see
Section~\ref{sec:algoDec}).  The problem is with the basic classes.
It follows directly from a theorem of Farrugia~\cite{farrugia:pnpc}
that deciding whether a graph can be partitioned into a triangle-free
part and a part that is disjoint union of cliques is NP-complete.  This
does not mean that recognizing ${\cal T}_1$ is NP-complete, because
one could take advantage of several features, such as being bull-free or
of the full definition of ${\cal T}_1$ in~\cite{bull2}.  We leave the
recognition of ${\cal T}_1$ as an open question.

\section{Extreme decompositions}
\label{sec:extDec}

The way we use decompositions for computing stable sets requires
building blocks of decomposition and asking at least two questions for
at least one block. When this process is recursively applied it
potentially leads to an exponential blow-up even when the
decomposition tree is linear in the size of the input trigraph.  This
problem is bypassed here by using what we call extreme decompositions,
that are decompositions whose one block of decomposition is basic and
therefore handled directly, without any recursive calls to the
algorithm.  In fact, some clever counting arguments might show that a
more direct approach leads to polynomially many questions, but we
consider extreme decompositions as interesting in their own right,
since they are very convenient to prove theorems by induction.  Hence,
we prefer to proceed as we do.

In this section, we prove that non-basic trigraphs in our
class actually have extreme decompositions.  We start by describing
the blocks of decomposition for the cutsets used in Theorem~\ref{thm:bull}.

We say that $(X, Y)$ is a \emph{decomposition} of a trigraph $T$ if
$(X, Y)$ is a partition of $V(T)$ and either $X$ is a homogeneous set of $T$,
or $X = A \cup B$ where $(A, B)$ is a small homogeneous pair or a
proper homogeneous pair of $T$.  The {\em block of decomposition
  w.r.t.\ $(X, Y)$ that corresponds to $X$}, denoted by $T_X$, is
defined as follows. If $X$ is a homogeneous set or a small homogeneous
pair, then $T_X=T[X]$. Otherwise, $X=A\cup B$ where $(A, B)$ is a
proper homogeneous pair, and $T_X$ consists of $T[X]$ together with
{\em marker vertices} $c$ and $d$ such that $c$ is strongly complete
to $A$, $d$ is strongly complete to $B$, $cd$ is a switchable pair,
and there are no other edges between $\{c,d\}$ and $A \cup B$.  The
{\em block of decomposition w.r.t.\ $(X, Y)$ that corresponds to $Y$},
denoted by $T_Y$, is defined as follows.  If $X$ is a homogeneous set,
then let $x$ be any vertex of $X$ and let $T_Y=T[Y \cup \{ x \}]$.  In
this case $x$ is called the {\em marker vertex} of $T_Y$.  Otherwise,
$X = A\cup B$ where $(A, B)$ is a homogeneous pair with split
$(A,B,C,D,E,F)$. In this case $T_Y$ consists of $T[Y]$ together with
two new \emph{marker vertices} $a$ and $b$ such that $a$ is strongly
complete to $C\cup E$, $b$ is strongly complete to $D\cup E$, $ab$ is
a switchable pair, and there are no other edges between $\{a,b\}$ and
$C \cup D \cup E \cup F$.

\begin{lemma}
  \label{e1}
  If $(X,Y)$ is a decomposition of a bull-free monogamous trigraph
  $T$, then the corresponding blocks $T_X$ and $T_Y$ are bull-free
  monogamous trigraphs.
\end{lemma}

\Proof 
Since all the edges in the blocks that go from marker vertices
to the rest of the block are strong edges, it follows that $T_X$ and
$T_Y$ are both monogamous trigraphs.

Suppose that $T_X$ or $T_Y$ contains a bull $H$. Since $H$ cannot be
isomorphic to an induced subtrigraph of $T$, it follows that $X=A \cup
B$ where $(A, B)$ is a homogeneous pair of $T$ and $H$ contains two
marker vertices from the block.  In a bull every pair of vertices has
a common neighbor or a common antineighbor. Since $c$ and $d$ do not
have a common neighbor nor a common antineighbor in $T_X$, it follows
that $H$ is a bull of $T_Y$ and $H$ contains $a$ and $b$. But then,
since $A$ is not strongly complete nor strongly anticomplete to $B$,
for some $a'\in A$ and $b'\in B$, $(V(H)\setminus \{ a,b\}) \cup \{
a',b'\}$ induces a bull in $T$, a contradiction.  \bbox

Let $(X,Y)$ be a decomposition of a trigraph $T$.  We say that $(X,
Y)$ is a {\em homogeneous cut} if $X$ is a homogeneous set or $X = A\cup B$
where $(A, B)$ is a proper homogeneous pair.  A homogeneous cut $(X,
Y)$ is {\em minimally-sided} if there is no homogeneous cut $(X', Y')$
with $X'\subsetneq X$.

\begin{lemma}\label{e2}
  If $(X, Y)$ is a minimally-sided homogeneous cut of a trigraph $T$,
  then the block of decomposition $T_X$, has no homogeneous cut.
\end{lemma}

\Proof
Assume not and let $(X', Y')$ be a homogeneous cut of $T_X$. 
We now consider the following two cases.

\vspace{2ex}

\noindent {\em Case 1:} $X$ is a homogeneous set of $T$.
\\
Since every vertex of $V(T)\setminus X$ is either strongly complete or
strongly anticomplete to $X$, it follows that $(X' , V(T)\setminus
X')$ is a homogeneous cut of $T$, contradicting our choice of $(X, Y)$
since $X' \subsetneq X$.
\\
\\
{\em Case 2:} $X= A \cup B$ where $(A, B)$ is a proper homogeneous
pair of $T$ with split $(A,B,C,D,E,F)$.
\\
Since $cd$ is a switchable pair of $T_X$, $\{ c,d\} \subseteq X'$ or
$\{ c,d\} \subseteq Y'$.

Suppose that $X'$ is a homogeneous set of $T_X$. Since $c$ and $d$ do
not have a common strong neighbour nor a common strong antineighbor,
it follows that $\{ c,d\} \subseteq Y'$.  Since $c$ is strongly
complete to $A$ and strongly anticomplete to $B$, $X'\subseteq A$ or
$X'\subseteq B$. But then $X'$ is a homogeneous set of $T$,
contradicting our choice of $(X, Y)$.

Therefore, $X'= A' \cup B'$ where $(A', B')$ is a proper homogeneous
pair of $T_X$ with split $(A', B', C', D', E', F')$.  First assume that $\{
c,d\} \subseteq Y'$.  Since $c$ is strongly complete to $A$ and
strongly anticomplete to $B$, it follows that $A' \subseteq A$ or
$A'\subseteq B$, and $B' \subseteq A$ or $B'\subseteq B$.  Hence
$(A',B')$ is a homogeneous pair of $T$.  We now obtain a contradiction
to the choice of $(X, Y)$ by showing that $(A',B')$ is in fact a
proper homogeneous pair of $T$.  If $A' \cup B'\subseteq A$, then $c
\in E'$, $d\in F'$ (i.e. $(C'\cup D')\cap \{ c,d \}=\emptyset$) and
hence, since $C'$ and $D'$ are nonempty, $(A',B')$ is a proper
homogenous pair of $T$.  So by symmetry we may assume that
$A'\subseteq A$ and $B'\subseteq B$. But then since $C$ and $D$ are
nonempty, and $C$ (resp. $D$) is strongly complete to $A$ (resp. $B$)
and strongly anticomplete to $B$ (resp. $A$), it follows that
$(A',B')$ is a proper homogeneous pair of $T$.

Now assume that $\{ c,d\} \subseteq X'$.  Since $C'$ and $D'$ are
nonempty, and no vertex of $T_X$ is strongly complete nor strongly
anticomplete to $\{ c,d \}$, we may assume w.l.o.g.\ that $c \in A'$
and $d \in B'$.  Hence $E'=F'=\emptyset$, $C'\subseteq A$ and
$D'\subseteq B$.  If $C'$ is strongly complete or strongly anticomplete 
to $D'$, then since $|C'\cup D'|\geq 3$, $C'$ or $D'$ is a
homogeneous set of $T_X$ and we obtain a contradiction as above. So we
may assume that $C'$ is not strongly complete nor strongly
anticomplete to $D'$.  But then, since $C$ and $D$ are nonempty,
$(C',D')$ is a proper homogeneous pair of $T$, contradicting our
choice of $(X, Y)$.  \bbox

\begin{theorem}\label{e3}
  Let $T$ be a bull-free monogamous trigraph that has a decomposition.
  If $T$ has a small homogeneous pair $(A, B)$, then let $X = A \cup
  B$ and $Y= V(T) \setminus X$.  Otherwise let $(X, Y)$ be
  minimally-sided homogeneous cut of $T$. Then the block of
  decomposition $T_X$ is basic.
\end{theorem}

\Proof
If $X = A \cup B$ where $(A, B)$ is a small homogenous pair then clearly $T_X\in {\cal T}_0$,
so assume that 
$T$ has no small homogeneous pair and that
$(X, Y)$ is a minimally-sided homogeneous cut of $T$.
By Lemma \ref{e2}, $T_X$ has no homogeneous cut. If $T_X$ has no small homogeneous pair,
then by Theorem~\ref{thm:bull} and Lemma \ref{e1}, 
 $T_X\in {\cal T}_0 \cup {\cal T}_1 \cup \overline{{\cal T}_1}$.
 So assume that $T_X$ has a small homogeneous pair with split
 $(A',B',C',D',E',F')$.  Set $X'= A' \cup B'$ and $Y' = V(T_X)
 \setminus X'$.  We now consider the following two cases.
\vspace{2ex}

\noindent {\em Case 1:} $X$ is a homogeneous set of $T$.
\\
Since every vertex of $V(T)\setminus X$ is either strongly complete or
strongly anticomplete to $X$, it follows that $(X',V(T)\setminus X')$ is
a small homogeneous pair of $T$, contradicting the assumption that $T$
has no small homogeneous pair.
\\
\\
{\em Case 2:} $X=A \cup B$ where $(A,B)$ is a proper homogeneous pair
of $T$ with split $(A,B,C,D,E,F)$.
\\
Since $cd$ is a switchable pair of $T_X$, $\{ c,d\} \subseteq X'$ or
$\{ c,d\} \subseteq Y'$.

First assume that $\{ c,d\} \subseteq Y'$.  Since $c$ is strongly
complete to $A$ and strongly anticomplete to $B$, it follows that $A'
\subseteq A$ or $A'\subseteq B$, and $B' \subseteq A$ or $B'\subseteq
B$.  Hence $(A',B')$ is a small homogeneous pair of $T$, contradicting
the assumption that $T$ has no small homogeneous pair.

Now assume that $\{ c,d\} \subseteq X'$. Since no vertex of $T_X$ is
strongly complete nor strongly anticomplete to $\{ c,d\}$ we may
assume w.l.o.g.\ that $c\in A'$ and $d\in B'$. Hence
$E'=F'=\emptyset$, $C'\subseteq A$ and $D'\subseteq B$.  But then,
since $C$ and $D$ are nonempty, either $(C',D')$ is a proper
homogeneous pair of $T$ or a subset of $C'\cup D'$ is a homogeneous
set of $T$ (if $C'$ is either strongly complete or strongly
anticomplete to $D'$), contradicting the minimality of $(X, Y)$.
\bbox

\section{Algorithms for finding decompositions}
\label{sec:algoDec}

The fastest known algorithm for finding a homogeous set in a graph is
linear time (see Habib and Paul~\cite{HabibP10}) and the fastest one
for the homogeneous pair runs in time $O(n^2m)$ (see Habib, Mamcarz,
and de~Montgolfier~\cite{HaMaMo:HP}).  But we cannot use these
algorithms safely here because we need minimally-sided decompositions
with several technical requirements (``small'', ``proper'') and we
need our algorithms to work for trigraphs.  However, it turns out that
all classical ideas work well in our context.

A $4$-tuple of vertices $(a, b, c, d)$ of a trigraph is \emph{proper}
if $ac$ and $bd$ are strong edges and $bc$ and $ad$ are strong
antiedges.  A proper 4-tuple $(a, b, c, d)$ is \emph{compatible} with
a homogeneous pair $(A, B)$ if $a\in A$, $b\in B$ and $c, d\notin
A\cup B$ (note that $c, d$ must be respectively in the sets $C, D$
from the definition of a split of a homogeneous pair).

\begin{lemma}\label{l:forcingHP}
  Let $T$ be a trigraph and $Z = (a, b, c, d)$ a proper $4$-tuple of
  $T$.  There is an $O(n^2)$ time algorithm that given a set $R_0
  \subseteq V(T)$ of size at least $3$ such that $Z \cap R_0 = \{a,
  b\}$, either outputs two sets $A$ and $B$ such that $(A, B)$ is a
  proper homogeneous pair of $T$ compatible with $Z$ and such that
  $R_0 \subseteq A\cup B$, or outputs the true statement ``There
  exists no proper homogeneous pair $(A, B)$ in $T$ compatible with
  $Z$ and such that $R_0 \subseteq A \cup B$''.

  Moreover, when $(A, B)$ is output, $A\cup B$ is minimal with respect to
  these properties, meaning that $A\cup B \subseteq A' \cup B'$ for
  every homogeneous pair $(A', B')$  satisfying the properties.
\end{lemma}

\Proof
We set $R= R_0$ and $S = V(T) \setminus R$, and we 
implement several forcing rules, stating that some sets of vertices
must be moved from $S$ to $R$.

We give mark $\alpha$ to all vertices of $V(T)$ that are strongly
adjacent to $c$ and strongly antiadjacent to $d$.  We give mark
$\beta$ to all vertices of $V(T)$ that are strongly adjacent to $d$
and strongly antiadjacent to $c$.  We give mark $\varepsilon$ to all
vertices of $V(T)$ not marked so far.  Observe that $a$, $b$, $c$ and $d$
receive marks $\alpha$, $\beta$, $\varepsilon$ and $\varepsilon$
respectively.

Vertices of $R$ should be thought of as ``vertices that must be in
$A\cup B$''.  Vertices with mark $\alpha$ should be thought of as
``vertices that are in $A$ if they are in $R$'';  vertices with mark
$\beta$ should be thought of as ``vertices that are in $B$ if they are in
$R$'';   and vertices with mark
$\varepsilon$ should be thought of as ``vertices that should not be in 
$R$'' . Note that the adjacency to $c$ and $d$ is enough to distinguish the three
cases, and this is why the marks are not changed during the process.

Here are the rules.  While there exists a  vertex $x\in R$ that is
marked, we apply them to $x$, and we unmark $x$. 

\begin{itemize}
\item If $x$ has mark $\varepsilon$, then stop and output ``There
  exists no homogeneous pair $(A, B)$ in $T$ compatible with $Z$ and
  such that $R_0 \subseteq A\cup B$''.
\item If $x$ has mark $\alpha$, then move the following sets from $S$
  to $R$: $\sigma(x)\cap S$, $(\eta(a) \cap S) \setminus \eta(x)$ and $(\eta(x)\cap
  S) \setminus \eta(a)$.
\item If $x$ has mark $\beta$, then move the following sets from $S$
  to $R$: $\sigma(x)\cap S$, $(\eta(b) \cap S) \setminus \eta(x)$ and $(\eta(x)\cap
  S) \setminus \eta(b)$.
\end{itemize}

If a vertex with mark $\varepsilon$ is in $R$, then no
homogeneous pair compatible with $(a, b, c, d)$ contains all vertices
of $R$; this explains the first rule.  If a vertex $x$ is in $R$, then
all switchable pairs with end $x$ must be entirely in $R$; this
explains why we move $\sigma(x)\cap S$ to $R$.  If a vertex $x$ in
$R$ has mark $\alpha$, it must share the same neighborhood in $S$ as
$a$; this explains the second rule.  The third rule is explained
similarly for vertices marked $\beta$.

The following properties are easily checked to be invariant during all
the execution of the procedure.  This means that they are true before
we start applying the rules, and they remain true after applying the
rules to each vertex.

\begin{itemize}
\item $R$ and $S$ form a partition of $V(T)$ and $R_0\subseteq  R$.

\item For all unmarked $v \in R$, and all $u\in S$, $uv$ is not a
  switchable pair.  

\item All unmarked vertices belonging to $R \cap \eta(c)$ have the
  same neighborhood in $S$, namely $S \cap \eta(a)$ (and it is a
  strong neighborhood).

\item All unmarked vertices belonging to $R\cap \eta(d) $ have
  the same neighborhood in $S$, namely $S \cap \eta(b)$ (and it is a strong
  neighborhood).

\item For every homogenous pair $(A, B)$ compatible with $(a, b, c,
  d)$ such that $R_0 \subseteq
  A\cup B$, we have  $R \subseteq A\cup B$ and
  $V(T)\setminus (A\cup B) \subseteq S$.
\end{itemize}

By the last item all moves from $S$ to $R$ are necessary.  This is why
the algorithm reports a failure if some vertex of $R$ has mark
$\varepsilon$.  If the process does not stop for that particular
reason, then all vertices of $R$ have been explored and are unmarked.
Note that $|R| \geq 3$ since $R_0\subseteq R$.  So, if $|S|\geq 3$ at the
end, we set $A= R\cap \eta(c)$, $B =R \cap \eta(d)$, and we observe
that $(A, B)$ is a proper homogeneous pair.

Since all moves from $S$ to $R$ are necessary, the homogeneous pair is
minimal as claimed.  This also implies that if $|S| < 3$, then no
proper homogeneous pair exists and we output this.  \bbox

\begin{lemma}\label{l:forcingHS}
  Let $T$ be a trigraph and $(a, b)$ a pair of vertices from $T$.
  There is an $O(n^2)$ time algorithm that given a set $R_0 \subseteq
  V(T)$ such that $a, b \in R_0$, either outputs a homogeneous set $X$
  such that $R_0 \subseteq X$, or outputs the true statement ``There
  exists no homogeneous set $X$ in $T$ such that $R_0 \subseteq X$''.

  Moreover, when $X$ is output, $X$ is minimal with respect to
  these properties, meaning that $X \subseteq X'$ for every
  homogeneous set $X'$ satisfying the properties.
\end{lemma}

\Proof The proof is similar to the previous one, so we just give a
sketch.  We mark all vertices except $a$ and we move $\sigma(a)$ to
$R$.  While there exists a marked vertex $x$ in $R$, we move
$\sigma(x)$, $\eta(x) \setminus \eta(a)$ and $\eta(a) \setminus
\eta(x)$ to $R$, and we unmark $x$.  \bbox

\begin{theorem}
  \label{th:algo}
  There exists an $O(n^8)$ time algorithm whose input is a trigraph
  $T$.  The output is a small homogeneous pair of $T$ if some exists.
  Otherwise, if $G$ has a homogeneous cut, then the output is a
  minimally-sided homogeneous cut.  Otherwise, the output is: ``$T$
  has no small homogeneous pair, no proper homogeneous pair and no
  homogeneous set''.
\end{theorem}

\Proof We search for a small homogeneous pair by enumerating all sets
of vertices of size at most $6$.  This can be done in time $O(n^8)$
($n^6$ for the enumeration, and $n^2$ to check wether a given small
set is a homogeneous pair).  If no small homogeneous pair is detected,
we first run the algorithm from Lemma~\ref{l:forcingHS} for all pairs
of vertices.  We then run the algorithm from Lemma~\ref{l:forcingHP}
for all proper $4$-tuples $(a, b, c, d)$ of $T$ and vertex $e$ with
$R_0 = \{a, b, e\}$.  Among the (possibly) outputted homogeneous sets
and pairs, we choose one of minimum cardinality.  This forms a
minimally-sided cut.  \bbox

\section{Computing $\alpha$ in \{bull, odd-hole\}-free graphs}
\label{sec:bofFree}

The maximum stable set problem is NP-hard for bull-free
graphs~\cite{poljak:74} and its complexity is not known for
odd-hole-free graphs.  In this section, we prove that it is polynomial
for the intersection of the two classes.

A graph $G$ is \emph{perfect} if every induced subgraph $H$ of $G$
satisfies $\chi(H) = \omega(H)$.  We use the following classical
results.

\begin{theorem}[Gr\"{o}tschel, Lov\'asz, and Schrijver~\cite{gls}]
  \label{th:gls}
  There is a polynomial time algorithm for the maximum stable set
  problem restricted to perfect graphs. 
\end{theorem}

\begin{theorem}[Chudnovsky, Robertson, Seymour and Thomas~\cite{crst}]
  \label{th:spgt}
  Every Berge graph is perfect. 
\end{theorem}

\begin{theorem}[Chudnovsky, Cornu\'ejols, Liu, Seymour, and
  Vu\v{s}kovi\'c~\cite{cclsv}]
  \label{th:reco}
  There is a polynomial time algorithm that decides whether an input
  graph is Berge. 
\end{theorem}

Observe that despite the previous result, the complexity of deciding
whether a graph contains an odd hole is not known.  We also need the
next classical algorithm that we use as a subroutine.  For faster
implementations (that we do not need here), see Makino and
Uno~\cite{makinoU04}.

\begin{theorem}[Tsukiyama, Ide, Ariyoshi, and Shirakawa
  \cite{tsukiyamaIAS77}]
  \label{th:enumS}
  There exists an algorithm for generating all maximal stable sets in
  a given graph $G$ that runs with $O(nm)$ time delay (i.e.\ the
  computation time between any consecutive output is bounded by
  $O(nm)$; and the first (resp.\ last) output occurs also in $O(nm)$
  time after start (resp.\ before halt) of the algorithm).
\end{theorem}

For the sake of induction, we need to work with weighted trigraphs.
Here, a \emph{weight} is a non-negative integer.  By a \emph{weighted
  trigraph with weight function $w$}, we mean a trigraph $T$ such that:

\begin{itemize}
\item every vertex $a$ has a weight  $w(a)$; 
\item every switchable pair $ab$ of $T$ has a weight $w(ab)$;
\item for every switchable pair $ab$, $\max \{ w(a), w(b)\} \leq w(ab)
  \leq w(a) + w(b)$.
\end{itemize}

Let $S$ be a stable set of $T$.  Recall that $\nu(T)$ denotes the set
of all strongly antiadjacent pairs of $T$, and $\sigma(T)$ the set of
all semiadjacent pairs of $T$. We set $c(S) = \{v\in S: \forall u\in
S\setminus \{ v\} , uv \in \nu(T)\}$.  We set $\sigma(S) = \{uv \in
\sigma(T): u, v \in S\}$.  Observe that if $T$ is monogamous, then for
every vertex $v$ of $S$, one and only one of the following outcomes is
true: $v\in c(S)$ or for some unique $w \in S$, $vw \in \sigma(S)$.
The \emph{weight of a stable set $S$} is the sum of the weights of the
vertices in $c(S)$ and of the weights of the (switchable) pairs in
$\sigma(S)$.  From here on, $T$ is a weighted monogamous
trigraph and $\alpha(T)$ denotes the maximum weight of a stable set of
$T$.

When $(X, Y)$ is a decomposition of $T$, we already defined the block
$T_Y$.  We now explain how to give weights to the marker vertices and
switchable pairs in $T_Y$.  Every vertex and switchable pair in $T[Y]$
keeps its weight.  If $X$ is a homogeneous set, then the marker vertex
$x$ receives weight $\alpha(T[X])$.  If $X = A \cup B$ where $(A, B)$
is a homogeneous pair, then we give weight $\alpha_A =\alpha(T[A])$ to
marker vertex $a$, $\alpha_B= \alpha(T[B])$ to marker vertex $b$ and
$\alpha_{AB}= \alpha(T[A\cup B])$ to the switchable pair $ab$.  It is
easy to check that the inequalities in the definition of a weighted
trigraph are satisfied. 

\begin{lemma}
  \label{l:pAlpha}
  $\alpha (T)=\alpha (T_{Y})$.
\end{lemma}
 
\Proof If $X$ is a homogeneous set, then this is clearly true since if
a maximum weight stable set $S$ of $T$ contains a vertex of $X$, then
$S\cap X$ is a maximum weight stable set of $T[X]$.

Suppose that $X = A \cup B$ where $(A, B)$ is a homogeneous pair with
split $(A, B, C, D, E, F)$.  Let $S$ be a maximum weighted stable set
of $T$.  If $S\cap (A\cup B)=\emptyset$, then $S$ is a stable set of
$T_{Y}$.  If $\emptyset \subsetneq S\cap (A\cup B)\subseteq A$, then
$S\cap A$ is a stable set of $T$ of weight $\alpha_A$, and hence
$(S\setminus A) \cup \{ a\}$ is a stable set of $T_{Y}$ of the same
weight as $S$.  If $\emptyset \subsetneq S\cap (A\cup B)\subseteq B$,
then $S\cap B$ is a stable set of $T$ of weight $\alpha_B$, and hence
$(S\setminus B) \cup \{ b\}$ is a stable set of $T_{Y}$ of the same
weight as $S$.  If $S\cap A\neq \emptyset$ and $S\cap B\neq
\emptyset$, then $S\cap (A\cup B)$ is a stable set of $T$ of weight
$\alpha_{AB}$, and hence $(S\setminus (A\cup B)) \cup \{ a,b\}$ is a
stable set of $T_{Y}$ of the same weight as $S$.  Therefore $\alpha
(T)\leq \alpha(T_{Y})$.  The reverse inequalities can be shown
similarly, and hence the result holds.  \bbox

\begin{lemma}
  \label{ohf1}
  If $(X,Y)$ is a decomposition of an odd-hole-free trigraph $T$, then
  the corresponding blocks $T_X$ and $T_Y$ are odd-hole-free.
\end{lemma}

\Proof Assume not and let $H$ be an odd hole contained in $T_X$ or
$T_Y$. Since $H$ cannot be isomorphic to an induced subtrigraph of
$T$, it follows that $X=A\cup B$ where $(A,B)$ is a homogeneous pair
of $T$ and $H$ contains two marker vertices from the block.  Note that
the two marker vertices either have a common neighbor or a common
antineighbor on $H$. Since $c$ and $d$ do not have a common neighbor
nor a common antineighbor in $T_X$, it follows that $H$ is an odd hole
of $T_Y$. But then, since $A$ is not strongly complete nor strongly
anticomplete to $B$, for some $a'\in A$ and $b'\in B$, $(V(H)\setminus
\{ a,b \} )\cup \{ a',b'\}$ induces an odd hole in $T$, a
contradiction.  \bbox

\begin{lemma}\label{ohf2}
  If $T$ is a trigraph from ${\cal T}_1$, then $T$ does not contain an
  antihole of length at least 7.
\end{lemma}

\Proof Let $T\in {\cal T}_1$ and let $X,K_1, \ldots ,K_t$ be
a partition of vertices of $T$ as in the definition of ${\cal
  T}_1$. Suppose that $H = h_1 \dots h_7\dots$ is an antihole of length at
least 7 in $T$.

In $H$, every 5-tuple of vertices contains a triangle (to see this,
start from a vertex of $H$ not in the 5-tuple, walk along $H$ and pick every
second vertex of the 5-tuple: they form a triangle).  Hence, $X$
contains at most four vertices of $H$, and $K_1 \cup \cdots \cup K_t$
contains at least three vertices of $H$. Since $H$ contains no strong
stable set of size~3, we may assume that $H \subseteq K_1 \cup K_2
\cup X$.  Suppose that  $K_1$ contains at least 3 vertices of $H$.  Then $K_2$
contains no vertices of $H$ (because in $H$, every triangle is
dominant).  No two vertices of $H\cap K_1$ are consecutive in $H$.  It
follows that $H \cap X$ contains a clique of the same size as $H\cap
K_1$, a contradiction (since $X$ contains no triangle).
So $K_1$, and similarly $K_2$ contains at most two vertices of $H$.
If $K_1$ and $K_2$ both contains two vertices of $H$, then the
complement of $H$ contains a 4-cycle, a contradiction.

It follows that we may assume that $H$ contains two vertices in $K_1$,
one in $K_2$ and four in $X$.  Furthemore, the vertices in $K_1\cup
K_2$ must be consecutive so that w.l.o.g.\ $h_1 \in K_1$, $h_2 \in K_2$, $h_3
\in K_1$ and $h_4, h_5, h_6, h_7 \in X$.  Hence, $N(h_2) \cap X$
contains a path on 4 vertices, so it does not partition into two
strong stable sets that are strongly complete to each other, a
contradiction.
\bbox

Let $T$ be a weighted monogamous trigraph with weight function $w$ and
a switchable pair $ab$.  We now define four ways to get rid of the
switchable pair $ab$ while keeping $\alpha$ the same.  This is needed
because sometimes we rely on algorithms for \emph{graphs}.  There are
four ways because $a$ (resp.\ $b$) can be transformed into a strong edge
or a strong antiedge.  Only one way is needed in this section, but in
Section~\ref{sec:polyOracle}, the four ways are needed.

The weighted monogamous trigraph $T_{a\rightarrow S}$ (resp.\
$T_{b\rightarrow S}$) is constructed as follows: replace switchable
pair $ab$ with a strong edge $ab$; add a new vertex $a'$ (resp.\ $b'$)
and make it strongly complete to $N_T(a)\setminus \{ b\}$ (resp.\
$N_T(b)\setminus \{ a\}$) and strongly anticomplete to the remaining
vertices; keep the weights of vertices and switchable pairs of
$T\setminus \{ a\}$ (resp.\ $T\setminus \{b\}$) the same; assign the
weight $w(a)+w(b)-w(ab)$ to $a$ (resp.\ $w(a)+w(b)-w(ab)$ to $b$) and
the weight $w(ab)-w(b)$ to $a'$ (resp.\ $w(ab) - w(a)$ to $b'$).  

The weighted monogamous trigraph $T_{a\rightarrow K}$ (resp.\
$T_{b\rightarrow K}$) is constructed as follows: replace switchable
pair $ab$ with a strong edge $ab$; add a new vertex $a'$ (resp.\ $b'$)
and make it strongly complete to $\{a\} \cup N_T(a) \setminus \{ b\}$ (resp.\
$\{b\} \cup N_T(b) \setminus \{ a\}$) and strongly anticomplete to the remaining
vertices; keep the weights of vertices and switchable pairs of
$T\setminus \{ a\}$ (resp.\ $T\setminus \{b\}$) the same; assign the
weight $w(a)$ to $a$ (resp.\ $w(b)$ to $b$) and
the weight $w(ab)-w(b)$ to $a'$ (resp.\ $w(ab) - w(a)$ to $b'$). 

Note that by the inequalities in the definition of a weighted
trigraph, all weights of vertices in $T_{a\rightarrow S}$,
$T_{b\rightarrow S}$, $T_{a\rightarrow K}$ and $T_{b\rightarrow K}$
are nonnegative.

\begin{lemma}\label{ohf3}
  If $T$ is a weighted monogamous trigraph and $ab$ is a switchable pair of
  $T$, then the following hold.
\begin{itemize}
\item[(i)] If $T$ is Berge then $T_{a\rightarrow S}$, $T_{b\rightarrow
    S}$, $T_{a\rightarrow K}$ and $T_{b\rightarrow K}$ are Berge.
\item[(ii)] $\alpha(T_{a\rightarrow S}) = \alpha(T_{b\rightarrow S}) =
  \alpha(T_{a\rightarrow K}) = \alpha(T_{b\rightarrow K}) = \alpha
  (T)$.
\end{itemize} 
\end{lemma}

\Proof We prove the statement for $T' = T_{a\rightarrow S}$, the other
proofs are similar.  To prove (i) assume $T$ is Berge, but $T'$
contains an odd hole or an odd antihole $H$.  Since $H$ cannot be
isomorphic to an induced subtrigraph of $T$, it must contain at least
two vertices of $\{ a',a,b\}$. If $H$ does not contain both $a$ and
$a'$, then by replacing the strong edge or strong antiedge of $H$ that
goes from $\{ a',a\}$ to $\{b\}$ by a switchable pair $ab$, we obtain
an odd hole or an odd antihole of $T$, a contradiction.  So $H$
contains both $a$ and $a'$. Observe that $a$ and $a'$ are not
contained in any switchable pair of $T'$. Since $H$ is of length at
least 5, it contains a vertex that is adjacent to $a'$ but not to $a$,
a contradiction.

To prove (ii), first let $S$ be a maximum weighted stable set of $T$.
If $S\cap \{ a,b\}=\{ a\}$ then let $S'=S\cup \{ a'\}$, if $S\cap \{
a,b\}=\{ a,b\}$ then let $S'=(S\setminus \{ a\})\cup \{ a'\}$, and
otherwise let $S'=S$. Then $S'$ is a stable set of $T'$ of the same
weight as the weight of $S$ in $T$, and hence $\alpha (T)\leq \alpha
(T')$.  Now let $S$ be a maximum weighted stable set of $T'$.  Note
that we may assume w.l.o.g. that $S\cap \{ a,a',b\} =\emptyset,
\{a,a'\},\{ b\}$ or $\{ a',b\}$.  If $S\cap \{ a,a',b\}=\{ a,a'\}$
then let $S'=S\setminus \{ a'\}$, if $S\cap \{ a,a',b\}=\{ a',b\}$
then let $S'=(S\setminus \{ a'\})\cup \{ a\}$, and otherwise let
$S'=S$. Then $S'$ is a stable set of $T$ of the same weight as the
weight of $S$ in $T'$, and hence $\alpha (T')\leq \alpha (T)$,
completing the proof of (ii).  \bbox

\begin{lemma}
  \label{l:countT1bar}
  If $T$ is a trigraph from $\overline{{\cal T}_1}$, then $T$ contains
  at most  $|V(T)|^3$ maximal stable sets.  
\end{lemma}

\Proof Consider sets $X, K_1, \ldots ,K_t$ that partition
$V(\overline{T})$ as in the definition of ${\cal T}_1$.  A maximal
stable set in $T$ is formed by a subset $S$ of size at most 2 of $X$
together with all the non-neighbors of $S$ in some $K_i$.  Therefore,
there are at most $n^3$ maximal stable sets in $T$.  \bbox

\begin{lemma}
  \label{algo:T1bar}
  There exists an $O(n^4m)$ time algorithm whose input is any trigraph
  $T$ and whose output is a maximum weighted stable set of $T$, or a
  certificate that $T$ is not in $\overline{{\cal T}_1}$.
\end{lemma}

\Proof Let $G$ be the realization of $T$ obtained by transforming
every switchable pair of $T$ by a non-edge.  Note that a subset of
$V(T) = V(G)$ is a stable set in $G$ if and only if it is a stable set
in $T$.  So, the problem of enumerating all maximal stable sets of $G$
is equivalent to the problem of enumerating all maximal stable sets of
$T$.  Note also that if $S$ is a stable set of $T$ and $S' \subseteq
S$, then $w(S') \leq w(S)$.

The algorithm uses Theorem~\ref{th:enumS} to enumerates all maximal
stable sets of $T$ (but stops if more than $n^3$ sets are found).
Lemma~\ref{l:countT1bar} certifies that if more than $n^3$ sets are
found, then $T$ is not in $\overline{{\cal T}_1}$.  Otherwise, among
all enumerated stable sets, the algorithm outputs one of maximum
weight.  \bbox

\begin{theorem}\label{ohf4}
There exists a polynomial-time algorithm with the following specifications.
\begin{description}
\item[Input:] A weighted monogamous trigraph $T$.
\item[Output:] Either $T$ is correctly identified as not being \{bull,
  odd-hole\}-free, or a maximum weighted stable set of $T$ is
  returned.
\end{description} 
\end{theorem}

\Proof We verify that $T$ is bull-free by checking all subsets of
vertices of size 5.  So let us assume that $T$ is bull-free.  We apply
the algorithm of Theorem~\ref{th:algo} to $T$.

Suppose first that no decomposition if found.  By
Theorem~\ref{thm:bull}, $T$ is basic.  If $T$ is in ${\cal T}_0$ (and
it is trivial to know whether $T$ is actually in ${\cal T}_0$), we
rely on some constant time brute force method.  If $T$ is not in
${\cal T}_0$, we run the algorithm from Lemma~\ref{algo:T1bar}.  So,
we have the maximum weighted stable set, or we know that $G$ is not in
$\overline{{\cal T}_1}$.  In this last case, we know that $T$ is in
${\cal T}_1$ and for every switchable pair $ab$ of $T$, we replace $T$
by $T_{a\rightarrow S}$, until we obtain a graph $G$.  We check
whether $G$ is Berge by Theorem~\ref{th:reco}.  If $G$ is Berge, then
$G$ is perfect by Theorem~\ref{th:spgt}, so we compute a maximum
weighted stable set of $G$ by Theorem~\ref{th:gls}, which is what we
need by Lemma~\ref{ohf3}.  Otherwise, $G$ is not Berge, so $T$ is not
Berge by Lemma~\ref{ohf3}.  Hence, $T$ contains an odd hole by
Lemma~\ref{ohf2}, so it is identified as not being \{bull,
odd-hole\}-free.

Suppose now that a decomposition $(X, Y)$ is found.  By
Theorem~\ref{e3}, $T_X$ is basic.  So, as shown in the paragraph
above, we may compute in polynomial time the maximum weight of a
stable set in $T_X$, or certify that $T_X$ is not in the class, but
then by Lemmas~\ref{ohf1} and~\ref{e1}, $T$ is identified as not being
in the class.  We can  also do this for  induced subtrigraphs of $T_X$.  Hence,
we can compute the weights needed to build $T_Y$.  We compute recursively
$\alpha(T_Y)$, that is equal to $\alpha(T)$ by Lemma~\ref{l:pAlpha}.
Since $T_Y$ has less vertices than~$T$, the number of recursive calls
is bounded by $|V(T)|$.  \bbox

Our algorithm relies on Gr\"{o}tschel, Lov\'asz, and
Schrijver's algorithm that colors perfect graphs~\cite{gls}. We wonder
whether a more direct approach exists. 

\begin{question}
  Is there a polynomial time combinatorial algorithm that computes
  $\alpha(G)$ for any input \{bull, odd hole\}-free graph $G$?
\end{question}

\section{Computing $\alpha$ in bull-free graphs}
\label{sec:decAlpha}

In this section, we use positive weights (no vertex nor switchable
pair in a trigraph has weight~0).  Also, switchable pairs have weight
at least~2.  

Let $R(x, y)$ be the smallest integer $n$ such that every graph on at
least $n$ vertices contains a clique of size $x$ or a stable set of size
$y$.  By a classical theorem of Ramsey, $R(3, x) \leq {x+1 \choose
  2}$.  We now define two functions $g$ and $f$ by $g(x) = {x+1
  \choose 2} - 1$ and $f(x) = g(x) + (x-1)({g(x) \choose 2} + 2g(x)
+1)$.  Note that $f(x) = O(x^5)$.  The next lemma handles basic
trigraphs.

\begin{lemma}
  \label{l:cBase}
  There exists an $O(n^4m)$-time algorithm with the following
  specifications.
  
  \begin{description}
  \item[Input:] A weighted monogamous basic trigraph $T$ on $n$
    vertices, in which all vertices have weight at least 1 and all
    switchable pairs have weight at least 2, with no homogeneous set,
    and a positive integer $W$.
  \item[Output:] One of the following true statements.
    \begin{enumerate}
    \item $n \leq f(W)$;
    \item the number of maximal stable sets in $T$ is at most $n^3$;
    \item  $\alpha(T) \geq W$.
    \end{enumerate}
  \end{description}
\end{lemma}

\Proof Let $G$ be the realization of $T$ in which all switchable
pairs are assigned value "strong antiedge". Note that $G$ is a graph.
We claim that testing whether output $i$ is true or not can be done in
polynomial time for $i=1, 2$.  For $i=1$, this is trivial and for
$i=2$, it follows from Theorem~\ref{th:enumS} applied to $G$.  The
algorithm does these two tests, stops if one of them is a success, and
if each attempt fails, it gives the answer~3.  The running time is
clearly $O(n^4m)$. It remains to check that when output~3 is the
answer it is a true statement.  So suppose for a contradiction that
$\alpha(T) < W$.  In particular, $W\geq 2$.

If $T$ is a trigraph in ${\cal T}_0$, then $n\leq 8 = f(2) \leq f(W)$,
so the algorithm should have stopped to give outcome~1, a
contradiction.  If $T$ is a trigraph in $\overline{{\cal T}_1}$, then
by Lemma~\ref{l:countT1bar}, the number of maximal stable sets in $T$
is at most $n^3$.  So, the algorithm should have stopped to give
outcome~2, a contradiction.

So, suppose that $T$ is a trigraph in ${\cal T}_1$, and consider the sets
$X, K_1, \ldots ,K_t$ as in the definition of ${\cal T}_1$.  If $|X|
\geq {W+1 \choose 2}$, then by Ramsey Theorem, $G$ contains a stable
set of size at least $W$, and therefore $T$ contains a stable set of
weight at least $W$ (since weights of vertices are at least 1 and
weights of switchable pairs are at least 2), a contradiction.  So,
$|X| \leq g(W)$.  If $t\geq W$, then by taking a vertex in each $K_i$,
$i= 1, \dots, t$, we obtain a stable set of size at least $W$, a
contradiction.  So $t \leq W-1$.

If for some $i \in \{1, \dots, t\}$ we have $|K_i| \geq {g(W) \choose
  2} + 2g(W) + 2$, then since $T$ is monogamous and $|X| \leq g(W)$,
at least ${g(W) \choose 2} + g(W) + 2$ vertices in $K_i$ are not
adjacent to any switchable pair and we call $K'_i$ the set formed by
these vertices (so, $|K'_i| \geq {g(W) \choose 2} + g(W) + 2$).
Consider the hypergraph $N$ with vertex set $X$ and hyperedge set
$\{N(v)\cap X|v\in K'_i\}$ and observe that $N$ has Vapnik-Cervonenkis
dimension bounded by 2 (for an introduction to Vapnik-Cervonenkis
dimension, see~\cite{alonSpencer:pm}).  Indeed, assume for
contradiction that $S=\{x_1, x_2, x_3\}$ is a shattered subset of
(three) vertices of $N$, i.e.\ for every subset $Y$ of $S$ there
exists a hyperedge $e$ of $N$ such that $S\cap e=Y$.  This would imply
the existence of three vertices $y_1,y_2,y_3$ in $K'_i$ such that
$y_i$ is joined only to $x_i$ in $S$, for $i=1,2,3$. Since $X$ is
triangle-free, there exists an antiedge in $S$, say $x_1x_2$.  But
then a contradiction appears since $\{y_1,y_2,y_3,x_1,x_2\}$ induces a
bull. Since the VC-dimension is at most 2, by Sauer's
Lemma~\cite{sauer}, the number of distinct hyperedges of $N$ is at
most ${|X|\choose 2}+|X|+1$, so at most ${g(W)\choose 2}+g(W)+1$.  But
since two distinct vertices of $K'_i$ have distinct neighborhoods to
avoid homogeneous sets, it follows that $K'_i$ has size bounded by
${g(W) \choose 2} + g(W) +1$, a contradiction.  So, $|K_i| \leq {g(W)
  \choose 2} + 2g(W) + 1$.

We proved that $|X| \leq g(W)$, $t\leq W - 1$ and for $i \in \{1,
\dots, t\}$, $|K_i| \leq {g(W) \choose 2} + 2g(W) + 1$.  It follows
that $$n \leq g(W) + (W-1)\left({g(W) \choose 2} + 2g(W) + 1\right) =
f(W).$$ So, the algorithm should have stopped to give outcome~1, a
contradiction.

\bbox

\begin{theorem}
  There is an algorithm with the following specification. 
  \begin{description}
  \item[Input:] A weighted monogamous bull-free trigraph $T$ and a positive
    integer $W$.
  \item[Output:] ``YES'' if $\alpha(T)\geq W$ and otherwise an
    independent set of maximum weight.
  \item[Running time:] $2^{O(W^5)}n^9$
    \end{description}
\end{theorem}

\Proof First, we delete all vertices of weight~0, and for all
switchable pairs of weight~1, we replace the switchable pair by a
strong edge.  It is easy to check that this does not change $\alpha$.
Now, all vertices have weight at least~1, and all switchable pairs
have weight at least~2.  Apply the algorithm from Theorem~\ref{th:algo}.

Suppose that no decomposition is found.  In particular, $T$ has no
homogeneous set.  Also by Theorem~\ref{thm:bull}, $T$ is basic.  Run
the algorithm from Lemma~\ref{l:cBase}.  If outcome~1 is the answer,
we compute by brute force a maximum weighted stable set in time
$2^{O(W^5)}$.  If outcome~2 is the answer, we compute a maximum
weighted stable set in polynomial time by Theorem~\ref{th:enumS}
applied to the realization of $T$ in which all switchable pairs are
assigned value ``strong antiedge''.  In both cases, we know the answer.
Finally, if outcome~3 is the answer, then we have that $\alpha(T) \geq
W$ and we output ``yes''.

Suppose that a decomposition $(X, Y)$ is found.  By Theorem~\ref{e3},
$T_X$ is basic.  We run the algorithm from Lemma~\ref{l:cBase} for
$T_X$.  If outcome~3 is the answer, output $\alpha(T)\geq W$, which is
the right answer since $\alpha(T[X]) \leq \alpha(T)$.  If outcome~1
or~2 is the answer, then compute a maximum weighted stable set in
$T[X]$ as above (if $X=A\cup B$ where $(A, B)$ is a homogeneous pair,
then we also compute a maximum weighted stable set in $T[A]$ and
$T[B]$ that are basic).  We now have the weights needed to construct
the block $T_Y$.  Run the algorithm recursively for $T_Y$ (this is
correct by Lemma~\ref{l:pAlpha}).  Since $T_Y$ has fewer vertices than
$T$, the number of recursive calls is bounded by $n$.  \bbox

\section{A polynomial Turing-kernel}
\label{sec:polyOracle}

Once an FPT-algorithm is found, the natural question is to ask for a
polynomial kernel for the problem. Precisely, is there a
polynomial-time algorithm which takes as input a bull-free
graph $G$ and a parameter $k$ and outputs a bull-free
graph $H$ with at most $O(k^c)$ vertices and some integer $k'$
such that $G$ has a stable set of size $k$ if and only if $H$ has a
stable set of size $k'$. Unfortunately, we have the following.

\begin{theorem}
  Unless NP $\subseteq$ coNP/poly, there is no polynomial kernel for
  the problem $\alpha(G)\geq k$, where $G$ is a bull-free graph and
  $k$ is the parameter.
\end{theorem}

\Proof This simply follows from the facts that the unparameterized
version of $\alpha(G)\geq k$ is NP-hard for bull-free graphs, and that
the problem is OR-compositional (see~\cite{bodl}). Indeed, if we are
given a family $G_1,\dots ,G_\ell$ of bull-free graphs and some
integer $k$, one can form the complete sum $G$ of these graphs by
taking disjoint copies of them and joining them pairwise by complete
bipartite graphs (i.e.\ for all $i\neq j$, put all edges between $G_i$
and $G_j$).  We then have that $G$ is bull-free, and moreover
$\alpha(G)\geq k$ if and only if there exists some $i$ for which
$\alpha(G_i)\geq k$ (this is the definition of an OR-compositional
problem).  By a result of Bodlaender et al.~\cite{bodl}, unless NP
$\subseteq$ coNP/poly, no NP-hard OR-compositional problem can admit a
polynomial kernel.  \bbox

Somewhat surprisingly, the non existence of a polynomial kernel is not
related to the hard core of the algorithm (computing the leaves) but
is related to the decomposition tree itself (since even complete sums
cannot be handled).  Indeed, our algorithm is a kind of kernelisation:
the answer is obtained in polynomial time provided that we compute a
stable set in a linear number of basic trigraphs of size at most $k^5$
(the leaves of our implicit decomposition tree).  A similar behaviour
was discovered by Fernau et al~\cite{Fernau} in the case of finding a
directed tree with at least $k$ leaves in a digraph ({\em Maximum Leaf
  Outbranching problem}): a polynomial kernel does not exist, but $n$
polynomial kernels can be found.  In our case, the leaves of the
decomposition tree are pairwise dependent, hence our method does not
provide $O(n^c)$ independent kernels of size $O(k^5)$. It seems that
the notion of kernel is not robust enough to capture this kind of
behaviour in which the computationally hard cases of the problem admit
polynomial kernels, but the (computationally easy) decomposition
structure does not.

Let $f$ be a computable function.  A parameterized problem has an {\em
  $f$-Turing-kernel} (see Lokshtanov~\cite{lokshtanov:phd}) if there
exists a constant $c$ such that computing the solution of any instance
$(X,k)$ can be done in $O(n^c)$ provided that we have unlimited access
to an oracle which can decide any instance $(X',k')$ where $(X', k')$
has size at most $f(k)$.

\begin{theorem}
  Stability in  bull-free weighted trigraphs (resp.\ graphs) has an
  $O(k^5)$-Turing-kernel.  The unweighted versions of both problems
  also have an  $O(k^5)$-Turing-kernel.
\end{theorem}

\begin{figure}
  \begin{center}
  \includegraphics{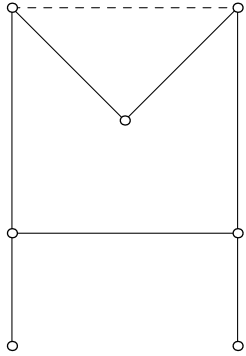}
  \end{center}
  \caption{A bull-free trigraph where all ways to expand a switchable
    pair creates a bull.\label{fig:badExp}}
\end{figure}

\Proof The proof is done already for weighted trigraphs. For weighted
graphs, there is a problem: with the present proof, we reduce graphs
to trigraphs, so we need to interpret a trigraph as a graph.  It is
not the case that every (integer) weighted bull-free trigraph can be
interpreted as an unweighted bull-free graph with the same $\alpha$.
Indeed, it is false in general that for every switchable pair $ab$ of
a bull-free trigraph, at least one of the trigraph $T_{a\rightarrow
  S}$, $T_{b\rightarrow S}$, $T_{a\rightarrow K}$ or $T_{b\rightarrow
  K}$ is bull-free.  In Fig.~\ref{fig:badExp}, we show an example of a
bull-free trigraph with a switchable pair represented by a dashed
line, where all the four obtained graphs contain a bull.  However, if
we start with a bull-free \emph{graph} and compute leaves of the
decomposition tree, every switchable pair in them is obtained at some
point by shrinking a homogeneous pair $(A, B)$ of a trigraph $T$ into
a switchable pair $ab$ of a trigraph $T'$.  Because of the requirement
that $A$ is not strongly complete and not strongly anticomplete to
$B$, we see that at least one of $T'_{a\rightarrow S}$,
$T'_{b\rightarrow S}$, $T'_{a\rightarrow K}$ or $T'_{b\rightarrow K}$
is in fact an induced subtrigraph of some semirealization of $T$ (and
recall that a trigraph is bull-free if and only if all its
semirealizations are bull-free).  By Lemma~\ref{ohf3}, this allows us
to represent the weighted bull-free trigraphs generated by our
Turing-kernel as bull-free graphs (with the same $\alpha$).

To prove the unweighted versions, just note that we can get rid of weights by
substituting a (strong) stable set on $w$ vertices for every vertex of weight~$w$.  \bbox

\section{Bounding $\chi$}
\label{sec:bChi}

A \emph{coloring} of a graph is an assignment of colors to the
vertices in such a way that no two adjacent vertices receive the same
color.  A \emph{semicoloring} of a graph is an assignment of colors
to the vertices in such a way that no maximal clique of $G$ is
monochromatic.  Clearly, every coloring is a semicoloring. Recall that
$\chi_T(G)$ is the triangle-free chromatic number of $G$, that is the
maximum chromatic number of a triangle-free induced subgraph of $G$. 

\begin{lemma}
  \label{l:semiChi}
  If $G$ is a bull-free graph,
  then $G$ admits a semicoloring with at most
  $$\max(\chi_T(G)+\omega(G), R(\omega(G) + 1, 3) - 1 + \omega(G))$$ colors.
\end{lemma} 

\Proof Our proof is by induction on $|V(G)|$.  We use
Theorem~\ref{thm:bull} and suppose first that $G$ is basic.  If $G$ is
in ${\cal T}_0$ (this is the base case of our induction), then clearly
it can be semicolored with eight colors (assign a different color to
each vertex).  If $G$ is in ${\cal T}_1$, its triangle-free part can
be colored with $\chi_T(G)$ colors and the cliques can be coloured
with $\omega(G)$ colors.  If $G$ is in $\overline{{\cal T}_1}$, then
we rely on Ramsey theory.  Let $X, K_1, \dots, K_i$ be the sets as in
the definition of ${\cal T}_1$.  Observe that $X$ contains no stable
set of size 3, and no clique of size $\omega(G) + 1$.  Therefore, by Ramsey Theorem, $X$ contains at
most $R(\omega(G) + 1, 3) - 1$ vertices and can be colored with
$R(\omega(G) + 1, 3) - 1$ colors.  Since $K_1 \cup \dots \cup K_i$
partitions into $i$ stable sets that are pairwise complete to one
another, it can  be colored with $\omega(G)$ colors.

Suppose now that $G$ admits a decomposition $(X, Y)$.  We may assume
that $G$ is connected.  The block $G_Y$ is bull-free by
Lemma~\ref{e1}, so every realisation $G'_Y$ of $G_Y$ has a coloring
with at most $\max(\chi_T(G)+\omega(G), R(\omega(G) + 1, 3) - 1 + \omega(G))$
colors (because $\omega(G_Y) \leq \omega(G)$).

If $X$ is a homogeneous set, then we color $G$ by giving to all
vertices of $Y$ the same color as in $G_Y$ and by giving to the
vertices of $X$ the color of the marker vertex $x$.  Since $G$ is
connected, no maximal clique of $G$ is contained in $X$, so the
coloring that we obtain is a semicoloring.

If $X= A\cup B$ where $A\cup B$ is a proper homogeneous pair with
split $(A, B, C, D, E, F)$, we consider the graph obtained from $G_Y$
by replacing the switchable pair $ab$ by an edge.  Observe that if $E
=\emptyset$, then $a$ and $b$ have different colors.  We color $G$ by
giving to all vertices of $Y$ the same color as in $G_Y$, to vertices
of $A$ the color of $a$ and to vertices of $B$ the color of $b$.  This
is a semicoloring of $G$, because no maximal clique of $G$ is included
in $A$ or in $B$, and one is included in $A\cup B$ only if
$E=\emptyset$.

If $X= A\cup B$ where $A\cup B$ is a small homogeneous pair with split
$(A, B, C, D, E, F)$, then we may assume that it is not proper, say
$C=\emptyset$.  If $E\neq \emptyset$, then the same proof as above
works, so suppose $E=\emptyset$.  Since $G$ is connected, we have
$D\neq \emptyset$.  We choose a vertex $b\in B$ and color by induction
the graph $G[Y \cup \{b\}]$.  We color $G$ by giving to all vertice of
$Y$ same the color that they have in this coloring and to vertices of
$B$ the color of $b$.  There are at most five vertices in $A$ and all
colors are available for them except the color of $b$.  We may
therefore color $A$ with at most five colors.  This is a semicoloring.
\bbox

\begin{theorem}
  There exists a function $f$ such that for every bull-free graph,
  $\chi(G) \leq f(\chi_T(G), \omega(G))$.  
\end{theorem}

\Proof By Lemma~\ref{l:semiChi}, there exists an increasing function
$g$ such that all bull-free graphs $G$ have a semicoloring with at
most $g(\chi_T, \omega(G))$ colors.  We set $f(x, y) = g(x, y) ^{g(x,
  y)}$.  We prove by induction on $\omega$ that every bull-free graph
$G$ with triangle-free chromatic number $\chi_T$ and maximum clique
size $\omega$ has a coloring with at most $f(\chi_T, \omega)$ colors.
If $\omega\leq 2$, this is clear because a semicoloring of a
triangle-free graph is a coloring.  Suppose $\omega>2$ and consider a
semicoloring of $G$ with $g(\chi_T, \omega)$ colors.  By considering
the color classes, we partition $G$ into $g(\chi_T, \omega)$ induced
subgraphs, and each of them has clique size at most $\omega-1$ and
triangle-free chromatic number at most $\chi_T$.
Therefore, by induction, we color $G$ with
$$g(\chi_T, \omega)f(\chi_T, \omega-1) = g(\chi_T, \omega) g(\chi_T,
\omega-1)^{g(\chi_T, \omega-1)}$$ $$\leq g(\chi_T, \omega) g(\chi_T,
\omega)^{g(\chi_T, \omega)-1} = f(\chi_T, \omega)$$ colors. \bbox

As observed in the introduction, the theorem above yields the
following. 

\begin{theorem}
  The class of \{bull, odd hole\}-free graphs is $\chi$-bounded. 
\end{theorem}

\section*{Acknowledgement}

Thanks to Andreas Brandst{\"a}dt, Maria Chudnovsky, Ignasi Sau and
Dieter Kratsch for several suggestions.  Thanks to Haiko M\"uller for
pointing out to us~\cite{farrugia:pnpc}.  Thanks to S\'ebastien
Tavenas and the participants to GROW 2013 for useful discussions on
Turing-kernels.

\end{document}